# High-precision nonlocal temporal correlation identification of entangled photon pairs for quantum clock synchronization

Runai Quan [1,2], Ruifang Dong [1,2,*], Xiao Xiang [1,2], Baihong Li[1,2,3], Tao Liu [1,2], Shougang Zhang [1,2]

[1] *Key Laboratory of Time and Frequency Primary Standards, National Time Service Center, Chinese Academy of Sciences, Xi'an, 710600, China*

[2] *School of Astronomy and Space Science, University of Chinese Academy of Sciences, Beijing, 100049, China*

[3] *College of Sciences, Xi'an University of Science and Technology, Xi'an, 710054, China*

*E-mail: dongruifang@ntsc.ac.cn

**Abstract:** High-precision nonlocal temporal correlation identification in the entangled photon pairs is critical to measure the time offset between remote independent time scales for many quantum information applications. The first nonlocal correlation identification was reported in 2009, which extracts the time offset via the algorithm of iterative fast Fourier transformations (FFTs) and their inverse. The best identification resolution is restricted by the peak identification threshold of the algorithm, and thus the time offset calculation precision is limited. In this paper, an improvement for the identification is presented both in the resolution and precision via a modified algorithm of direct cross correlation extraction. A flexible resolution down to 1 ps is realized, which is only dependent on the Least Significant Bit (LSB) resolution of the time-tagging device. The attainable precision is shown to be mainly determined by the inherent timing jitter of the single photon detectors, the acquired pair rate and acquisition time, and a sub picosecond precision (0.72 ps) has been achieved at an acquisition time of 4.5 s. This high-precision nonlocal measurement realization provides a solid foundation for the field applications of entanglement-based quantum clock synchronization, ranging and communications.

**Introduction**

Quantum correlated photon pairs generated from spontaneously parametric down conversion (SPDC) have been proven to behave superior performance. Owing to their strong temporal correlation (about a few hundred femtoseconds) in correlated photon pairs, the temporal coincidence identification has become an important facility in applications of quantum spectroscopy [1-4], quantum communication [5-8], and quantum clock synchronization [9-13], etc. Especially in case of clock synchronization applications utilizing energy-time entangled photon pair sources, an accurate time-offset measurement between remote sites via the coincidence identification is the prerequisite for high-precision synchronization. Typically, the temporal coincidence identification is implemented by a dedicated coincidence hardware, such as a time-correlated single photon counting device (TCSPC), which demands local detection of the correlated photons. In field applications, the photon pairs are usually distributed remotely and their time offset detection can only be implemented nonlocally. Therefore, the nonlocal coincidence identification of the time-energy entangled photon pairs is essential to the high-accuracy applications.

In 2009, Ho et al. proposed a first nonlocal time correlation identification algorithm [14]. In the algorithm, the recorded time sequences are firstly binned for subsequent fast Fourier transformations (FFTs). By implementing direct product on the Fourier transformed bins and applying the inverse FFTs on the outputs, the temporal coincidence identification can be nonlocally implemented between the photon pairs. The time offset addressed by the two clocks thus can be nonlocally extracted by finding the coincidence peak. As the time offset can be extracted exactly only when the identified coincidence peak exceeds a reference threshold determined by the bin size, the ultimate calculation resolution and the consequent precision are restricted.

In this paper, we present a modified algorithm of direct cross correlation extraction for the nonlocal coincidence identification. Due to the single-photon characteristics of the photon pair source, the recorded time sequences follow a sparse population. Thus, the cross correlation can be readily executed by counting the subtractions of the time sequences that are within a certain time window, which corresponds to the identification resolution. A coarse extraction is firstly launched onto a segment of the time sequences to determine the rough peak position of the time offset. Referenced by this coarse time offset, the coincidence distribution is then constructed by dealing with the counts of the subtractions that are within a much finer time window than the Glauber second-order correlation width of the detected photon pairs. Based on the algorithm, a flexible resolution down to 1 ps is realized, which is only determined by the Least Significant Bit (LSB) resolution of the utilized event timer (ET). The attainable precision as a function of the measurement time has further been analyzed, and a minimum precision of 0.72 ps at 4.5 s has been achieved. The result also shows that the attainable precision based on this modified algorithm is mainly determined by the detection setup, such as inherent timing jitter of the single photon detectors and the acquired pair rate, while has nothing to do with the data processing hardware.

**Algorithm description**

According to quantum theory, the joint detection probability of the time-correlated photon pair that are separately distributed to the space-time points $(\mathbf{r}_1, t_1)$ and $(\mathbf{r}_2, t_2)$ is proportional to the Glauber second-order correlation function [15, 16],

$$G^{(2)}(\mathbf{r}_1, t_1, \mathbf{r}_2, t_2) = \langle E^{(-)}(\mathbf{r}_1, t_1) E^{(-)}(\mathbf{r}_2, t_2) E^{(+)}(\mathbf{r}_2, t_2) E^{(+)}(\mathbf{r}_1, t_1) \rangle, \quad (1)$$

where $E^{(-)}$ and $E^{(+)}$ are the negative- and positive-frequency part of the electric field operators at space-time point $(\mathbf{r}_j, t_j), j = 1, 2$. In the stationary case, the second-order correlation function $G^{(2)}$ only depends on $t_1 - t_2$. By knowing the joint distribution and thus the peak position of the $G^{(2)}$ function, the time offset between the two space-time points is obtained.

In practical experiment, such distribution cannot be directly measured. Instead the coincidence counting rate $R_c$ within a certain time window is measured, which can be expressed as [17]

$$R_c \sim \int_0^T dt_1 dt_2\, S(t_1 - t_2 - t_0) G^{(2)}(t_1 - t_2), \quad (2)$$

where $T$ represents the data acquisition time. $S(t_1 - t_2 - t_0)$ is the coincidence window function centered at $t_0$, and can be given by a rectangular function

$$S(t_{1,i} - t_{2,j} - t_0) = \begin{cases} 1, |t_{1,i} - t_{2,j} - t_0| \leq \frac{\tau_{BW}}{2}; \\ 0, |t_{1,i} - t_{2,j} - t_0| > \frac{\tau_{BW}}{2}. \end{cases} \quad (3)$$

where $\tau_{BW}$ denotes the resolution of the coincidence measurement. When $\tau_{BW}$ is chosen to be so small that $S(t_1 - t_2 - t_0)$ is equivalent to a delta function, the probability of $G^{(2)}(t_1 - t_2)$ at $t_0$ is obtained. As the detected time events are actually discrete, the coincidence counting rate $R_c$ should be rewritten as

$$R_c(t_1 - t_2 = t_0) \sim \sum_{i=1}^{n} \sum_{j=1}^{m} S(t_{1,i} - t_{2,j} - t_0)\, G^{(2)}(t_{1,i} - t_{2,j}). \quad (4)$$

Where $n$ and $m$ are the numbers of the time sequences recorded for the signal and idler photons within the acquisition time of $T$, denoted as $\{t_{1,i}\}, i = 1, \cdots n$, and $\{t_{2,j}\}, j = 1, \cdots m$, respectively. Therefore, the issue of extracting the time offset is turned to search for the maximum coincidences from the two detected time sequences by varying $t_0$ with a step of $\tau_{BW}$.

Consider that the $G^{(2)}$ function has a FWHM width of $\Delta\tau$. By first setting the searching resolution about three times of $\Delta\tau$, i.e. $\tau_{BW,C} \approx 3\Delta\tau$, a quick identification procedure is applied to seek for the coarse position of $t_{0c,max}$ at which the coincidence rate reaches maximum. During this procedure, only a short segment of time sequences is used with a length of $M \ll n, m$. Assume the detected time events for the two space-time points $(\mathbf{r}_1, t_1)$ and $(\mathbf{r}_2, t_2)$ are referenced to a common clock, $t_{1,1} - t_{1,M} \approx t_{2,1} - t_{2,M}$. The value of $t_{0c}$ is then varied from $t_{0c,ini} = t_{1,1} - t_{2,1}$ to $t_{0c,fin} = t_{0,ini} + N_c\tau_{BW,C}$, with the integer $N_c \approx \frac{t_{1,M}-t_{1,1}}{\tau_{BW,C}} \approx \frac{t_{2,M}-t_{2,1}}{\tau_{BW,C}}$. Thus by successively applying the coincidence window function $S(t_{1,i} - t_{2,j} - t_{0c,k})$, where $t_{0c,k} = t_{0c,ini} + k\tau_{BW,C}$, to each $t_{1,i} - t_{2,j}$ with $i, j \in M$ and $k \in N_c$, a coarse coincidence identification with respect to $t_{0c,k}$ can be obtained. By finding the maximum coincidence counts, a coarse time offset $t_{0c,max}$ can be determined.

Centered at the coarsely identified time offset $t_{0c,max}$, a fine coincidence identification is subsequently executed. In this procedure, the variation of $t_{0f}$ is within the range of $t_{0c,max} \pm \tau_{BW,C}/2$ with a step of $\tau_{BW,f} \approx \tau_{BW,C}/N_f$, where $N_f$ is an integer and can be flexibly chosen as long as it is larger than 100. Then applying the similar procedure with the fine resolutions $\tau_{BW,f}$ leads to a final coincidence distribution as a function of $t_{0f,k'} = t_{0c,max} - \tau_{BW,C}/2 + k'\tau_{BW,f}$, $k' \in N_f$. Via a Gaussian fitting of the coincidence distribution, the expected time offset $\bar{\tau}$ corresponding to the maximum coincidences is therefore derived.

**Experimental implementation**

To demonstrate this nonlocal temporal correlation identification algorithm, a frequency-anticorrelated photon pair source was utilized. Fig. 1 shows the schematic diagram of the setup. The frequency anti-correlated biphoton source was generated by using a continuous wave 780-nm laser to pump a 10 mm-long type-II PPKTP crystal. The details of the experimental system can be found in Ref. [18]. After filtering out the residual pump, the orthogonally polarized signal and idler photons were departed by a fiber polarization beam splitter. The signal and idler photons were transmitted directly to the single photon detectors D1 and D2 (superconductive nanowire single-photon detectors, SNSPDs), with a quantum efficiency of 50% and a timing jitter of 70 ps [19, 20].

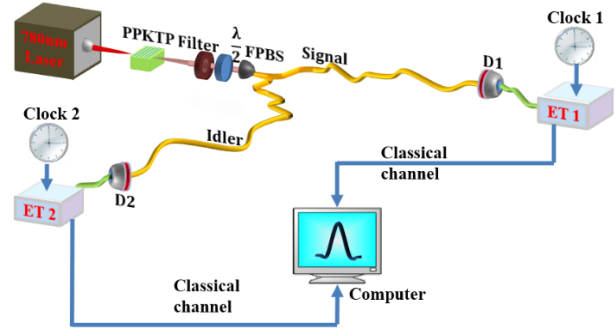

Fig. 1 Schematic experimental setup

The arrival times of the signal and idler photons to the detectors were recorded independently by two commercial event timers, ET 1 and ET 2 (A033-ET, Eventech Ltd), which are referenced to a laboratory-own H-maser frequency standard and a Rb atomic clock (PRS10, SRS), respectively. The recorded time sequences are tagged as $\{t_{1,i}\}$ and $\{t_{2,j}\}$. Based on the above algorithm, the time offset between the signal and idler photons at D1 and D2 are extracted.

With a photon pair acquisition rate of 6000 kcps, the coarse coincidence distribution of the entangled photon events acquired in $T_a = 0.14$ s was firstly constructed with the coarse resolution set to be $\tau_{BW,C} = 500\ ps$ and shown in Fig. 2(a). For convenience, the absolute time offset between two paths has been taken out from the results. It can be seen that, the coarse position of the coincidence peak can be distinctly distinguished, which has a height about 30 and significantly higher than that of the average background coincidence.

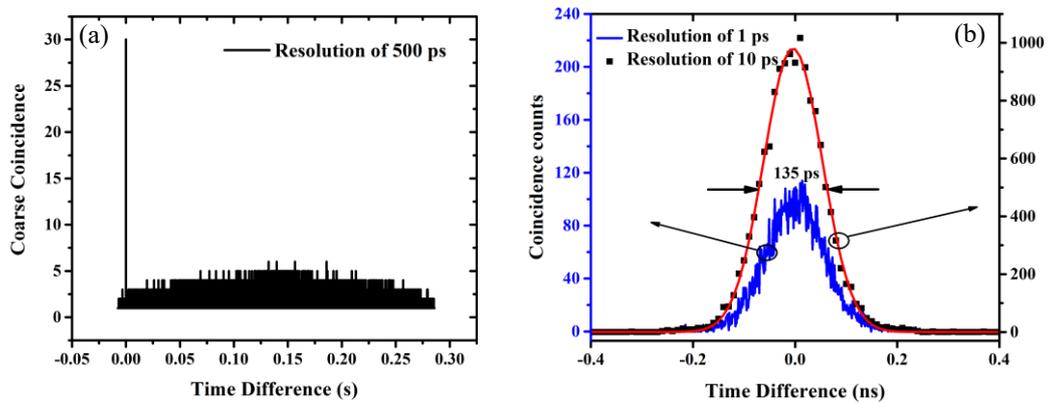

Fig. 2 (a)The extracted coarse coincidence distribution from the time tagged sequences acquired in 0.14 s, the resolution is set as 500 ps, and (b) the constructed coincidence distributions of entangled photons with resolutions of 10 ps (in black squares) and 1 ps (in blue solid line);

Centered at the coarsely identified time offset $t_{0c,max}$, the fine coincidence identification was subsequently executed. Based on the time sequences acquired in a time of $T_a = 2.2$ s, the coincidence distributions of the entangled photons, with the fine resolution set as $\tau_{BW,f} = 10$ ps (black squares) and 1 ps (blue solid line) respectively, are depicted in Fig. 2(b). For clarity, the coincidence curve with 1 ps resolution is rescaled by adjusting the range of Y-axis. Through Gaussian fitting, it can be seen that both distributions show a FWHM width about $\Delta\tau_{obs} \sim 135$ ps, which should be attributed to the contributions from both the SNSPD jitter and the relative frequency drift between the two atomic clock references [21, 22].

In order to further evaluate the precision performance of this nonlocal coincidence identification setup based on our modified algorithm, we further implement the measurement with a common reference clock. Based on the time sequences acquired in a time of $T_a = 4.5$ s, the constructed coincidence histograms are shown in Fig. 3 (a) by choosing the fine resolution ($\tau_{BW,f}$) successively as 1 ps (by black squares), 3 ps (by red circles), 5 ps (by blue up-triangles), 7 ps (by magenta down-triangles), 9 ps (by olive diamonds), 15 ps (by dark cyan stars), 25 ps (by navy left-triangles), 35 ps (by violet right-triangles) and 55 ps (by purple hexagons). Through Gaussian fitting, the FWHMs ($\Delta\tau_{obs}$) of these histograms at each resolution are acquired accordingly and given in Table 1. It can be seen that, as long as the chosen resolution is much smaller than the timing jitter of the SNSPD, i.e., $\tau_{BW,f} \ll \Delta\tau_{jitter}$, the measured coincidence FWHMs are in very good consistency and show the dominant contribution from the timing jitter of the SNSPD. Such consistency also shows that the minimum resolution can reach down to 1 ps for the coincidence identification. It should be noted that the achievable least resolution is determined by the Least Significant Bit (LSB) resolution of the time-tagging device. With the resolution further increased, the fitted FWHM tends to become larger than the nominal jitter, indicating an unsuitable setting of the coincidence identification resolution.

Table 1. The measured coincidence FWHMs in accordance with different fine resolutions.

| $\tau_{BW,f}$ (ps) | 1 | 3 | 5 | 7 | 9 | 15 | 25 | 35 | 55 |
|---|---|---|---|---|---|---|---|---|---|
| $\Delta\tau_{obs}$ (ps) | 70.49 | 70.67 | 70.93 | 70.83 | 71.03 | 71.57 | 72.62 | 75.99 | 80.87 |

The attainable precision in terms of standard deviation (SD) as a function of the chosen resolution is shown in Fig. 3 (b) by black dots. Similarly, the precision is approximately independent of the chosen resolution only when it is obviously smaller than the timing jitter of the SNSPD, i.e., $\tau_{BW,f} \ll \Delta\tau_{jitter}$. When $\tau_{BW,f}$ approaches or even surpass $\Delta\tau_{jitter}$, the precision tends to degrade dramatically. This infers that the timing jitter of the detector setup determines the least precision of the time offset measurement.

Based on the above Gaussian fittings, the extracted time offsets ($\bar{\tau}$) of the coincidence histograms acquired at each resolution are shown in Fig. 3 (b) as well by blue squares. One can see that, the value of $\bar{\tau}$ is proportional to the increment of the chosen resolution. It can be readily understood as $\bar{\tau}$ corresponds to the position with maximum coincidences binned by the chosen resolution. With a quadratic polynomial fitting, one can see a very good agreement with the results and the resolution-independent time offset is given by 140.32 ps, which is slightly smaller than the value of 140.82 ps acquired at the resolution of 1 ps. Therefore, besides the requirement of $\tau_{BW,f} < \Delta\tau_{jitter}$, a finer $\tau_{BW,f}$ is favorable for both constructing the coincidence distribution and optimizing the time offset identifications. Furthermore, for the application of high-precision quantum metrologies, a more accurate time offset measurement determines the ultimate performance of the system. However, in order to get a reliable coincidence, a longer data acquisition time is demanded for the coincidence calculation with a finer coincidence resolution.

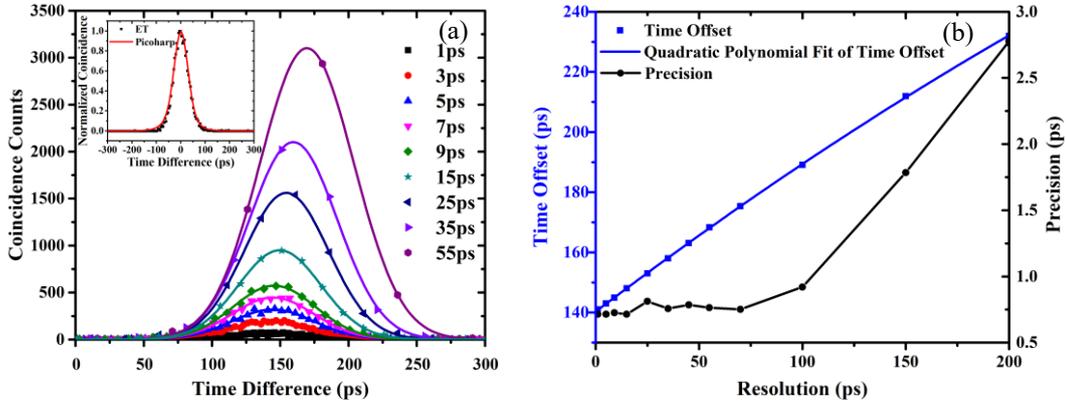

Fig. 3 (a) The constructed coincidence distributions of entangled photons acquired in 4.5s, with different resolutions of 1 ps (in black squares), 3 ps (in red circles), 5 ps (in blue up-triangles), 7 ps(in magenta down-triangles), 9 ps (in olive diamonds), 15 ps (in dark cyan stars), 25 ps (in navy left-triangles), 35 ps (in violet right-triangles) and 55 ps (in purple hexagons); and (b) the achieved precisions (in black circles), and time offsets (in blue squares) of the measured coincidences as a function of the resolution applied to the fine time offset identification, where each time offset value is acquired by the Gaussian fitting of the peak coincidence distribution at each resolution.

By fixing the resolution as $\tau_{BW,f}$=7 ps, the precision of the measured time offset was further investigated as a function of the acquisition time $T_a$, and the results are shown in Fig. 4 by black squares. It can be seen that, the precision is enhanced with the increase of $T_a$.

According to Ref. [21], the precision dependence on $T_a$ should follow

$$\delta t_{SD} = \frac{\Delta \tau'}{\sqrt{2 R_c T_a}}, \tag{5}$$

where $\Delta\tau'$ denotes the correlation width of the entangled photons together with the contribution from timing jitter of the detector setup, i.e., $\Delta\tau' = \sqrt{\Delta\tau^2 + \Delta\tau_{jitter}^2}$. In the experiment, $\Delta\tau' \approx \Delta\tau_{jitter}$. $R_c$ is the acquired coincidence count rate, which is about 1140 cps. Based on Eq. (5), the theoretical simulation is made and plotted in Fig. 4 (red solid line) as well. The agreement between theory and experimental results is shown at an acquisition time within 1 s. Afterwards, the dropping of the precision curve starts to deviate from the $1/\sqrt{2R_c T_a}$ slope, and a minimum value of 0.72 ps is finally achieved at $T_a$=4.5 s. The comparison between the acquired precision and theoretical simulation also reveals that the attainable precision based on this algorithm has reached the fundamental limit set by the detection setup.

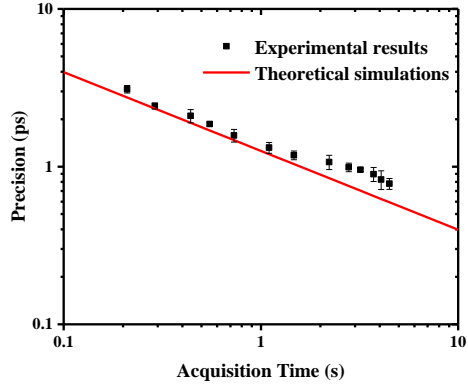

Fig. 4 The acquired precision of the measured time offsets versus the data acquisition time (in black squares), and the corresponding simulation (in red solid line).

**Discussion**

To virtually implement a remote identification of the coincidence, we further inserted a 10-km single-mode fiber in the signal path to realize the physical separation between signal and idler photons. The nonlocal coincidence identification was performed based on the recorded time sequences acquired at a time of 5 s and an average pair rate of 620 cps, and the constructed contribution is shown in Fig. 5 (a). One can see that, a resolution down to 1 ps can be applied to distinctly construct the coincidence distribution, though its shape is not smooth enough. For comparison, the algorithm proposed by Ho et al. [14] was also applied to the same time sequences for the coincidence identification. According to reference [14], the identified correlation peak is represented by the so-called statistical significance S, which is defined as the ratio between the correlation peak height above the baseline and the standard deviation of the latter. Limited by the computer memory, the maximum available bin size for FFT is chosen as $2^{27}$. Fig. 5 (b)-(d) depict the identified statistical significance S at a resolution of 9 ps, 4.5 ps and 2.25 ps, respectively. It can be seen that the correlation peak cannot be identified with a resolution of 2.25 ps. Such result can be explained by the calculated maximum correlation significance $S_{max}$ with respect to a threshold $S_p$, which represents the lowest peak reference for a fixed FFT bin size. Only when $S_{max}$ exceeds the threshold $S_p$, the correlation peak can be correctly identified. In our case, the $S_p$ is about 6.4 for $2^{27}$ FFT bins. Corresponding to the resolution of 9 ps and 4.5 ps, the calculated maximum correlation significance $S_{max}$ were 14.4 and 8.7 respectively, which are still larger than $S_p$. When the resolution was set to be 2.25 ps, as the calculated $S_{max}$ was 6.6 and smaller than $S_p$, the coincidence peak cannot be distinguished any more.

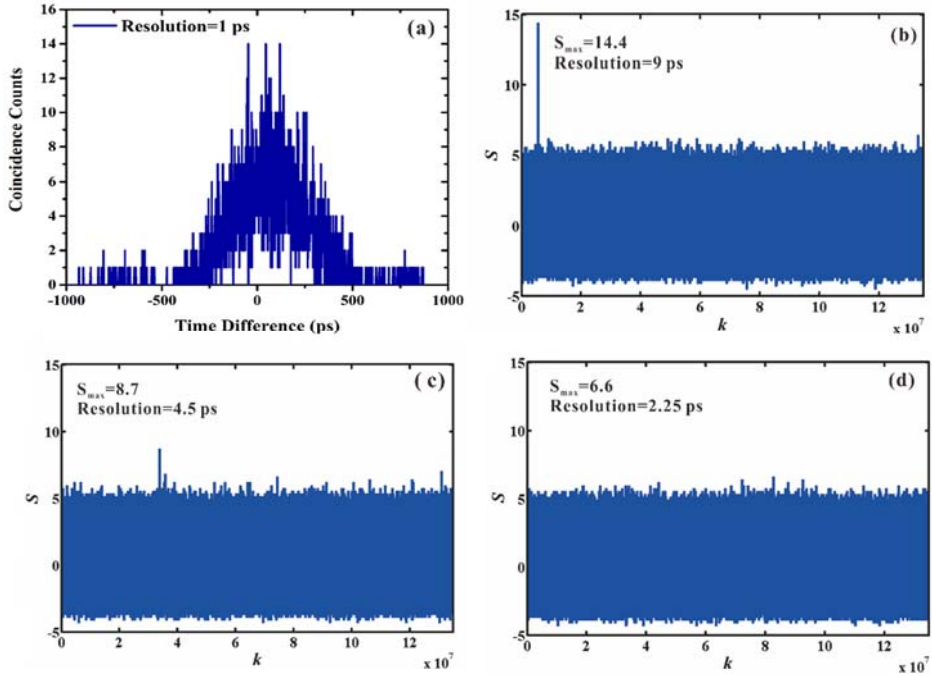

Fig. 5 The plots of identified coincidence between two photons separated by 10 km fiber links by utilizing our modified algorithm with a resolution of 1ps (a) and Ho's algorithm with the resolution set as 9 ps (b), 4.5 ps (c) and 2.25 ps (d). In Fig. (b)-(d), $S$ denotes the statistical significance of the time correlation results, and $k$ represents the correlation index, and the time offset can be obtained by the product of $k$ and the resolution.

In comparison with the dedicated coincidence hardware, the coincidence measurement can be achieved without deteriorating the resolution, meanwhile there is no restriction on the locations of the two correlated photons after propagation. Utilizing this algorithm, we have successfully demonstrated experimental realizations of the femtosecond-level quantum clock synchronization (QCS) [23] and the quantum nonlocality test based on nonlocal dispersion cancellation (NDC) [24] in laboratory.

**Conclusion**

In summary, we have demonstrated an improved algorithm for the nonlocal coincidence identification of entangled photons by using independent event timers. With this algorithm, we have implemented a time offset measurement between two independent atomic clocks based on a H-maser and a Rb oscillator, and a measurement resolution down to 1 ps, which is restricted by the LSB of the event timers, has been realized. Further using a common clock as the reference of both event timers, the attainable precision has been analyzed. According to the very good agreements between the theoretical simulation and the measured results, it has been shown that the attainable precision is mainly determined by the inherent timing jitter of the single photon detectors, the acquired pair rate and acquisition time. In the experiment, at a pair rate of 1140 cps and an acquisition time of 4.5s, a precision of 0.72 ps has been achieved. By increasing the acquired pair rate and acquisition time, the precision can be improved and a femtosecond precision can be expected. High-precision quantum clock synchronization and nonlocal dispersion cancellation have been successfully demonstrated benefitted from the algorithm used [23, 24]. Through the work, further progress can be expected in practical applications of field quantum metrologies and quantum communications and so on.

**Acknowledgements**

This work was supported by National Natural Science Foundation of China (12033007, 61801458, 61875205, and 91836301), "Western Young Scholar" Project of CAS (XAB2019B17, XAB2019B15), Key R&D Program of Guangdong province (2018B030325001), Frontier Science Key Research Project of CAS (QYZDB-SSW-SLH007), Strategic Priority Research Program of CAS (XDC07020200), Chinese Academy of Sciences Key Project (ZDRW-KT-2019-1-0103).

**Data Availability Statement**

The data that support the findings of this study are available from the corresponding author upon reasonable request.